\documentstyle[prl,aps,twocolumn,graphicx,floats]{revtex}

\begin{document}

\twocolumn[\hsize\textwidth\columnwidth\hsize\csname@twocolumnfalse\endcsname

\title
{Local Magnetic Order vs.\ Superconductivity in a Layered Cuprate}

\author{N. Ichikawa,$^{1*}$ S. Uchida,$^1$ J. M. Tranquada,$^2$ 
T. Niem\"oller,$^3$ P. M. Gehring,$^4$ S.-H. Lee,$^{4,5}$ and J. R.
Schneider$^3$}
\address{$^1$Department of Superconductivity, School of Engineering,
University of Tokyo,\\ 2-11-16 Yayoi, Bunkyo-ku, Tokyo 113-8656, Japan}
\address{$^2$Physics Department, Brookhaven National Laboratory,
Upton, NY 11973-5000}
\address{$^3$ Hamburger Synchrotronstrahlungslabor HASYLAB
 at Deutsches Elektronen-Synchrotron DESY,\\ Notkestr.\ 85, D-22603 Hamburg,
Germany}
\address{$^4$NIST Center for Neutron Research, National Institute of Standards
and Technology, Gaithersburg, MD 20899}
\address{$^5$University of Maryland, College Park, MD 20742}
\date{\today}
\maketitle 

\begin{abstract}
We report on the phase diagram for charge-stripe order in
La$_{1.6-x}$Nd$_{0.4}$Sr$_x$CuO$_4$, determined by neutron and x-ray
scattering studies and resistivity measurements.  From an analysis of the
in-plane resistivity motivated by recent nuclear-quadrupole-resonance
studies, we conclude that the transition temperature for local charge
ordering decreases monotonically with $x$, and hence that local
antiferromagnetic order is uniquely correlated with the anomalous depression
of superconductivity at $x\approx\frac18$.  This result is consistent with
theories in which superconductivity depends on the existence of charge-stripe
correlations.
\end{abstract}
]

Superconductivity in the layered cuprates is induced by doping charge
carriers into an antiferromagnetic insulator.  The kinetic energy of the
mobile carriers competes with the superexchange interaction between
neighboring Cu spins \cite{ande97,emer99}.   There is increasing evidence for
the hole-doped cuprates that this competition drives a spatial segregation of
holes which form antiphase domain walls between strips of
antiferromagnetically correlated Cu spins
\cite{niem99,hunt99,sing99,waki99,mook98,mook99}.  A major controversy
surrounds the issue of whether the mesoscopic self-organization of charges and
spins is a necessary precursor for high-temperature superconductivity
\cite{emer97,cast97,vojt99}, or whether it is simply an alternative
instability that competes with superconductivity \cite{rice97,bask99}.

To gain further insight into this problem, we have performed a systematic
study of the phase diagram of La$_{1.6-x}$Nd$_{0.4}$Sr$_x$CuO$_4$ (LNSCO), a
system for which evidence of competition between superconductivity and stripe
order has been reported previously \cite{tran97a}.  From neutron and x-ray
scattering measurements we show that the charge and magnetic ordering
temperatures reach their maxima at $x\approx\frac18$.  For $x<\frac18$, the
charge-ordering transition is limited by a structural phase boundary.  The
low-temperature structural phase involves a change in the tilt pattern of the
CuO$_6$ octahedra, stabilized by the substituted Nd, which can pin vertical
charge stripes \cite{tran95a}.  

At first glance, these results, together with the anomalous depression of the
superconducting transition temperature, $T_c$, at $x\approx\frac18$, appear to
provide confirmation that charge-stripe order is in direct competition with
superconductivity; however, the picture becomes more complicated when one
takes into account recent nuclear-quadrupole-resonance (NQR) studies of LNSCO
\cite{hunt99,sing99}.  In this work, a transition (involving
the onset of an apparent loss of intensity) has been identified which
coincides with the charge ordering determined by diffraction for
$x\ge\frac18$; however, in contrast to the diffraction results, the NQR
transition temperature, $T_{\rm NQR}$, continues to increase as $x$ decreases
below $\frac18$.  Furthermore, the same transition is observed in
La$_{2-x}$Sr$_x$CuO$_4$ (LSCO) for $x\lesssim\frac18$.  The implication is
that local charge order, not easily detected by diffraction techniques, may
occur even in the absence of the Nd-stabilized lattice modulation.

Does $T_{\rm NQR}$ really correspond to charge ordering?  To test this
possibility, we have analyzed the in-plane resistivity,
$\rho_{ab}(T)$, which should be a sensitive measure of charge
ordering.  Through a scaling analysis, we have identified a temperature scale
$T_u$ that corresponds to a low-temperature upturn with respect to an
extrapolated linear variation with $T$.  (The signature of the charge
ordering is subtle, as befitting its unconventional nature.)   We show that
$T_u$ corresponds with $T_{\rm NQR}$ for both LNSCO and LSCO,
thus providing support for the association of $T_{\rm NQR}$ with local charge
ordering. Furthermore, we show that $T_u$ and $T_{\rm NQR}$ are linearly
correlated with the size of the lattice distortion at low temperature. 
Together with the reasonable assumption that the magnitude of the
charge-order parameter at low temperature is correlated with the ordering
temperature, this result is strong evidence for a monotonic decrease of the
charge-order parameter with increasing hole concentration (over the range
studied here).

A monotonic variation of the stripe pinning strength means that there is no
correlation with the anomalous depression of $T_c$ at $x\approx\frac18$.  We
are left with the suprising conclusion that it is, instead, the static
magnetic order alone which has a special association with the $\frac18$
anomaly.  In making this assertion, we do not argue that ordering the charge
is good for superconductivity; to the contrary, $T_c$ is certainly reduced in
all of our LNSCO samples compared to comparably-doped LSCO.  Rather, our
point is that, while pinning charge stripes is not good, it is magnetic order
that is truly incompatible with superconductivity. 
The competition between static local antiferromagnetism
and superconductivity is supported by recent theoretical work \cite{havi99},
and is compatible with the spin-gap proximity-effect mechanism for
superconductivity \cite{emer97}.

For this study, a series of crystals of
La$_{2-x-y}$Nd$_{y}$Sr$_x$CuO$_4$, with $y=0.4$ and $x=0.08$ to 0.25, was
grown by the travelling-solvent floating-zone method \cite{ichi99}.   Figure
1(a) shows the electrical resistivity measured parallel to
the CuO$_2$ planes by the six-probe method.  As previously
reported \cite{tran96b}, there are upturns in $\rho_{ab}$ at low
temperature for the $x=0.12$ and 0.15 samples, compositions at which charge
order has been observed \cite{tran96b,tran95a,vonz98,niem99}.  In each there
is also a small jump near 70~K, where a subtle structural transition takes
place from the so-called low-temperature-orthorhombic (LTO) phase to the
low-temperature-tetragonal (LTT) or an intervening
low-temperature-less-orthorhombic (LTLO) phase \cite{axe89,craw91}.  At
$x=0.12$, charge ordering and the structural transition are essentially
coincident \cite{tran96b,vonz98}; however, charge ordering occurs
significantly below the structural phase change at $x=0.15$ (see Fig.~2)
\cite{niem99}.

The resistivity for $x=0.10$ looks somewhat different.  Instead of an
increase at the structural transition temperature, $\rho_{ab}$ {\it decreases}
below the transition, and continues to decrease in a typically metallic
fashion until superconductivity sets in.  To test whether stripe order occurs
in this sample, we performed a neutron scattering experiment at the NIST
Center for Neutron Research (NCNR) \cite{NIST}.  We found that the
$x=0.10$ sample does indeed exhibit charge and spin order. The temperature
dependence of the peak intensities for representative superlattice peaks are
shown in Fig.~1(b).  On warming, the charge order (which has also been
confirmed by x-ray diffraction measurements at HASYLAB) seems to be limited
by the structural transition at 65~K, while the magnetic order disappears at
a lower temperature.  

We have also used neutron scattering to determine the magnetic ordering
temperatures ($T_m$) in samples with $x=0.08$ and 0.25.  The results are
summarized in Fig.~2.  (Further details of the neutron studies will be
presented elsewhere.)  The new results for $x=0.08$ and 0.10 make it clear
that the highest $T_m$ occurs at
$x\sim\frac18$, where the superconducting transition ($T_c$) is most greatly
depressed.  Also plotted in the figure are the transition temperatures
($T_{\rm NQR}$) deduced from Cu NQR measurements by
Singer, Hunt, and Imai \cite{sing99,hunt99}.  Those temperatures coincide with
the charge-order transitions, $T_{ch}$, for $x=0.12$ and 0.15 determined by
diffraction, but there appears to be a discrepancy for $x<0.12$. 

\begin{figure}[t]
\centerline{\includegraphics[width=2.45in]{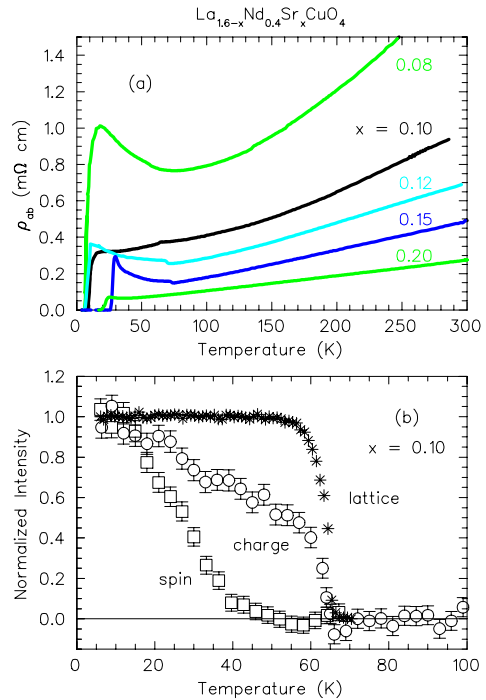}}
\medskip
\caption{(a) In-plane resistivity vs.\ temperature measured on single crystals
of La$_{1.6-x}$Nd$_{0.4}$Sr$_x$CuO$_4$ with several different Sr
concentrations.  (b) Neutron diffraction results for $x=0.10$.  Stars:
intensity of the (100) superlattice peak, which is allowed in the LTT and
LTLO phases, but not in LTO.  Circles: charge-order superlattice peak at
wave vector $(2+2\epsilon,0,0.5)$.  Squares: magnetic superlattice peak
at $(\frac12+\epsilon,\frac12,0)$. (Peak indexing is based on the simple
tetragonal unit cell \protect\cite{tran95a}.) In each case, peak intensity was
measured by scanning $T$ without moving the spectrometer. 
$T$-dependence of the background was measured and subtracted, and a
$T$-independent offset was applied.}
\end{figure}

The NQR and diffraction results for $x<0.12$ are not necessarily in
conflict, since NQR is an inherently local probe, whereas the diffraction
measurements require substantial spatial correlations of the charge order in
order to obtain detectable superstructure peaks.  But it is also interesting
that NQR measurements \cite{sing99,hunt99} suggest charge order in pure
La$_{2-x}$Sr$_x$CuO$_4$ for $x\lesssim0.125$, where diffraction studies have
not yet detected any charge-related superlattice peaks.  If some form of
charge ordering is occurring within the LTO phase, one would expect to see an
indication of it in the resistivity.  As we will show below, it is, in fact,
possible to identify a signature of charge order in resistivity measurements.

\begin{figure}[t]
\centerline{\includegraphics[width=2.45in]{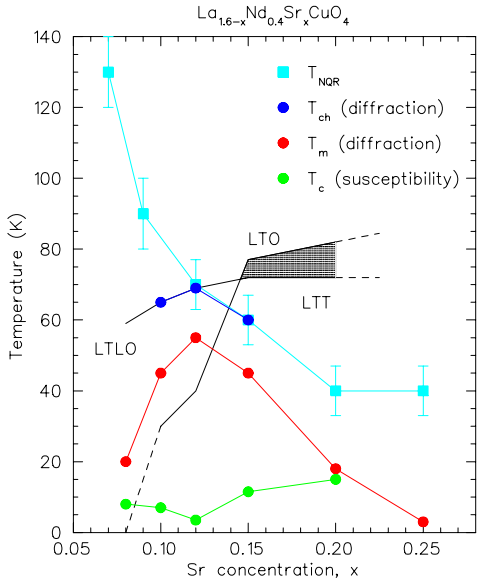}}
\medskip
\caption{(color) Phase diagram for
La$_{1.6-x}$Nd$_{0.4}$Sr$_x$CuO$_4$.\protect\linebreak  Light-blue squares:
$T_{\rm NQR}$ \protect\cite{sing99}; dark-blue circles:
$T_{ch}$ from diffraction studies \protect\cite{tran96b,vonz98,niem99} (and
present work); red circles: $T_m$ from neutron diffraction
\protect\cite{ichi99,tran97a}; green circles: $T_c$ from magnetic
susceptibility \protect\cite{ichi99,tran97a,oste97}. Lines through symbols are
guides to the eye.  Black lines indicate structural phase boundaries
determined by neutron diffraction \protect\cite{ichi99,tran96b}.
Shaded region indicates coexistence of LTO and LTT phases. The $x=0.25$
crystal appears to be a mixture of LTO and LTT phases with no obvious
transition between 10 and 300~K.}
\end{figure}

To analyze the resistivity, we consider first the behavior at
higher temperatures.  For cuprates doped to give the maximum $T_c$, it was
noted early on \cite{gurv87} that, over a surprisingly large temperature
range,
\begin{equation}
  \rho(T) = \alpha T + \beta,
\end{equation}
with $\beta$ very close to zero.  We find that this formula describes fairly
well the results in Fig.~1(a) for $T\gtrsim200$~K.  Values for $\alpha$ were
obtained by fitting Eq.~(1), with $\beta\equiv0$, to data in the range
$250\mbox{~K}<T<300$~K; the same analysis was also applied to resistivity
data for La$_{2-x}$Sr$_x$CuO$_4$ crystals with $x=0.10$, 0.12, 0.15, and 0.20
\cite{ichi99}.  

\begin{figure}[t]
\centerline{\includegraphics[angle=90,width=3.0in]{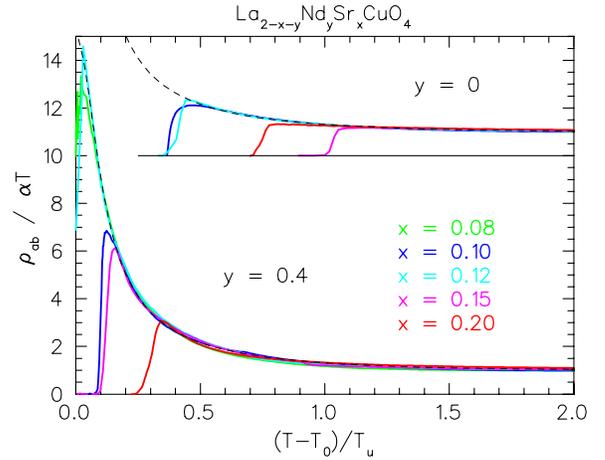}}
\medskip
\caption{(color) $\rho_{ab}$ divided by $\alpha T$ vs.\ $(T-T_0)/T_u$ for
La$_{2-x-y}$Nd$_y$Sr$_x$CuO$_4$ with $y=0.4$ and $y=0$ (shifted vertically). 
The values of $T_0$ and $T_u$, adjusted to scale the data sets onto the same
curve, are plotted in Fig.~5.  The black dashed line is a model function
described in the text.}
\end{figure}

Next, we analyze the upturn in the resistivity at low temperature.  The
temperature at which the upturn becomes significant varies with $x$, and so
does the rate of upturn; it was pointed out previously by B\"uchner and
coworkers \cite{huck98} that the rate of upturn increases monotonically as one
goes from $x=0.10$ to 0.12 to 0.15.  We have found that all of the data can be
scaled approximately onto a single curve if $\rho_{ab}$ is divided by 
$\alpha T$ and then plotted against a reduced temperature $t=(T-T_0)/T_u$,
where $T_u$ is the characteristic upturn temperature and $T_0$ is the
temperature towards which $\rho_{ab}$ appears to be diverging.  The scaled
resistivities are shown in Fig.~3; note that the same scaling is useful for
samples both with and without Nd.  The scaled curve is given approximately by
\begin{equation}
  \rho_{ab}/\alpha T = \tanh(15t)/\tanh(t),
\end{equation}
and we have determined error bars for the parameters $T_0$ and $T_u$ by
performing least-squares fits to this function.

The values of $T_u$ are compared with $T_{\rm NQR}$ in Fig.~4,
where both are plotted vs.\ the maximum orthorhombic splitting 
$(b-a)_{\rm LTO}$ in the LTO phase.  B\"uchner {\it et al.} \cite{buch94}
have shown that $(b-a)_{\rm LTO}$ is a useful measure of the octahedral tilt
angle, which changes orientation but not magnitude in the LTLO and LTT phases.
For the $y=0.4$ samples, we used our own neutron measurements of 
$(b-a)_{\rm LTO}$, while we used results from \cite{rada94} for LSCO.

\begin{figure}[t]
\centerline{\includegraphics[width=3.0in]{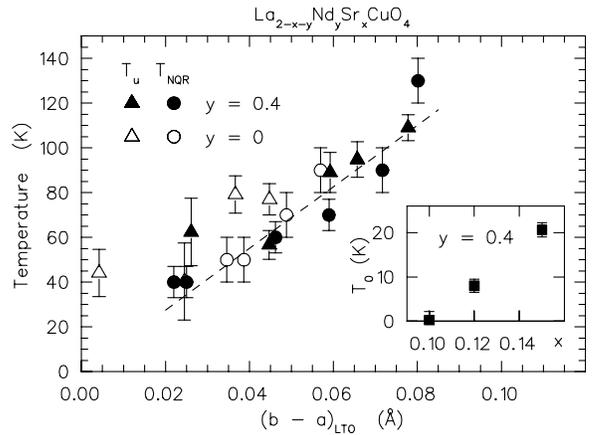}}
\medskip
\caption{Values of parameters $T_u$ (triangles) and $T_{\rm NQR}$ (circles)
as a function of $(b-a)_{\rm LTO}$ for La$_{2-x-y}$Nd$_y$Sr$_x$CuO$_4$
(filled symbols, $y=0.4$; open, $y=0$).  Dashed line is a guide to the eye. 
Inset: $T_0$ vs.\ $x$ for $y=0.4$.}
\end{figure}

From Fig.~4 we see that (1) the values of $T_u$ and $T_{\rm NQR}$ agree
within the error bars, and (2) both values tend to scale with the octahedral
tilt angle, independent of the tilt orientation (LTO vs. LTT).  The first
point reinforces the association of $T_{\rm NQR}$ with charge order, while
the second indicates that the ordering temperature for local charge ordering
is controlled by the tilt angle.  (A correlation between tilt angle and $T_c$
reduction was noted previously by Dabrowski {\it et al.} \cite{dabr96}.) 
Longer-range charge correlations (those detected by diffraction) appear to be
sensitive to the tilt orientation.

The variation of $T_0$ with $x$ is shown in the inset of Fig.~4 for
$y=0.4$. There is a considerable increase in $T_0$ from $x=0.10$ to 0.15.  We
suggest that this trend may be associated with a phase locking of
charge-density-wave correlations along neighboring charge stripes, a
possibility suggested by Kivelson, Fradkin, and Emery \cite{kive98}.  Whether
or not this interpretation is correct, there is clearly no correlation
between the variations of $T_0$ (or $T_u$) and the depression of $T_c$ for
$y=0.4$, which is greatest at $x\approx\frac18$.

The variations of $T_u$ and $T_0$ shown in Fig.~4 strongly indicate that
ordering of the charge stripes is not responsible for the strong depression
of $T_c$ at $x\approx\frac18$.  We are then left with the conclusion that the
culprit must be the magnetic order, which is maximized at the point where
$T_c$ is minimized.  That local antiferromagnetic order competes with
superconductivity is certainly compatible with the spin-gap proximity-effect
mechanism for superconductivity \cite{emer97}.  In that theory, hole pairing
is associated with the occurrence of a spin gap; given that
antiferromagnetic order competes with singlet correlations and a spin gap,
one would then expect $T_c$ to be depressed when magnetic order is present. 
(Of course, charge order is a prerequisite for magnetic order.) The trade off
between local magnetic order and superconductivity is also emphasized in a
recent numerical study \cite{havi99}.

One simple reason why $T_m$ might reach a maximum at
$x=\frac18$ is suggested by recent analyses of coupled spin ladders
\cite{twor99,kim99}.  If the charge stripes are rather narrow and centered on
rows of Cu atoms, then the intervening magnetic strips would consist of 3-leg
spin ladders.  Theoretical analyses have shown that even weak
couplings between a series of 3-leg ladders will lead to order at
sufficiently low temperature, whereas weakly coupled 2- or 4-leg ladders have
a quantum-disordered ground state \cite{twor99,kim99}.  As $x$ deviates from
$\frac18$, one would have a combination of even-leg and 3-leg ladders, thus
weakening the tendency to order.  Although there is no direct experimental
evidence concerning the registry of the stripes with the lattice, the picture
of a CuO$_2$ plane broken into a series of 3-leg ladders by Cu-centered charge
stripes at $x=\frac18$ is appealing in the present case.

One might argue that only longer-range magnetic (or charge) order is relevant
for suppressing superconductivity.  We believe that a counter-example is
given by the case of Zn-doping, where a local suppression of
superconductivity is associated with static short-range antiferromagnetic
correlations about the Zn sites \cite{juli00}.

In conclusion, we have presented evidence that it is local {\it magnetic}
order rather than charge-stripe order which is responsible for the anomalous
suppression of superconductivity in LNSCO at $x\approx\frac18$. 
While pinning charge stripes also causes some reduction of $T_c$, charge
order appears to be compatible with superconductivity as long as the spin
correlations remain purely dynamic.

This research was supported by the U.S.-Japan Cooperative Research Program on
Neutron Scattering, a COE Grant from the Ministry of Education, Japan, and
U.S. Department of Energy Contract No.\ DE-AC02-98CH10886.   We acknowledge
the support of the NIST, U.S. Department of Commerce, in providing the neutron
facilities used in this work; SPINS is supported by the National Science
Foundation under Agreement No.\ DMR-9423101.  NI and JMT acknowledge the
hospitality of the NCNR staff.  We thank V. J. Emery and S. A. Kivelson for
helpful comments.


\begin{thebibliography}{10}
\bibitem{NIST}
The $x=0.10$ sample was studied at spectrometer BT4 with 14.7-meV
  neutrons; $x=0.08$ and 0.25 samples were characterized at the NG-5 SPINS
  spectrometer with 5-meV neutrons.
\end{thebibliography}

\end{document}